%% file: main.tex
\pdfoutput=1
\documentclass[runningheads]{llncs}
\usepackage[T1]{fontenc}
\usepackage{graphicx}
\input{preamble}

\begin{document}
\title{Automated Abstraction Refinement for Information Flow Security in Embedded Systems\thanks{This preprint has not undergone peer review or any post-submission improvements or corrections. This contribution has been accepted for publication at the 24th International Conference on Software Engineering and Formal Methods (SEFM 2026); the Version of Record will appear in Springer's \emph{Lecture Notes in Computer Science} series and a link to it will be added here upon publication.}}
\titlerunning{Automated Abstraction Refinement for Information Flow Security in ES}
%
\author{Jonas Becker-Kupczok\orcidID{0009-0007-9605-006X} \and
Lukas Ernst\orcidID{0009-0002-9394-1821} \and
Paula Herber\orcidID{0000-0002-5349-154X}}
\authorrunning{J. Becker-Kupczok, L. Ernst, P. Herber}
%
\institute{University of Münster, Computer Science Department, Münster, Germany \\
\email{\{jonas.becker-kupczok,lernst,paula.herber\}@uni-muenster.de}}
\maketitle              
\begin{abstract}
Information flow analysis (IFA) is a powerful technique for verifying confidentiality and integrity and is therefore
highly desirable for security-sensitive embedded systems.
However, as these systems are inherently concurrent and time-dependent, 
existing IFA for embedded systems tend to be either imprecise or expensive.
In this paper, we propose an approach to tackle this problem using automatic abstraction refinement. The key idea is to heuristically choose abstraction levels based on information about dependencies between states and detected potential information leakage. Our approach builds on previous work, where we leverage symbolic execution 
to precisely capture data, control, timing, and event dependencies between processes within an IFA.
To capture values symbolically, this analysis uses abstract interpretation.
While the existing approach requires manual definition of abstraction levels, our novel contribution in this paper is using carefully designed heuristics to select these levels automatically.
The aim is to keep analysis times acceptable
while also retaining enough information to decide whether or not illegal information flow is possible.
We have implemented our approach for the system design language SystemC and demonstrate its feasibility with experimental results on several shared bus architectures.

\end{abstract}
\section{Introduction}

Embedded systems are ubiquitous and widely used in safety- and security-critical applications, \eg in cars or pacemakers. This makes it crucial to protect them against malevolent attacks.
One way of ensuring confidentiality or integrity is performing an information flow analysis (IFA) to verify non-interference between high (i.e. confidential or untrusted) inputs and low (i.e. public or trusted) outputs. However, embedded systems are typically time-dependent and concurrent, which makes the analysis challenging.
Existing IFA approaches usually suffer from one of two drawbacks: either they abstract completely from the concrete timing behavior, causing a severe loss of precision, or they precisely capture time- and control dependencies, which then means that suitable abstractions are required to enable scalable analysis.
These drawbacks are particularly severe in systems with shared bus architectures: IFA that ignore timing behavior always classify the whole system as insecure, whereas IFA that precisely capture time- and control dependencies, on the other hand, will consider all possible interleavings of read and write accesses to the bus, which does not scale well.

In previous work \cite{Becker-Kupczok2024a}, we have presented an approach that uses symbolic execution to compute an overapproximation of all possible interleavings in timed and concurrent systems.  To tackle the scalability problem, 
it combines symbolic execution 
with abstract interpretation, abstracting selected data variables while keeping track of the state of most processes, events, and other scheduling-relevant information. The results
are used 
to construct a dependence graph, which captures an overapproximation of possible information flow in the system.
The exact choice of the abstraction has a significant impact on the analysis: Keeping track of certain variables can reduce the state space, \eg by determining which branch is taken in an if-statement, but can also greatly increase it. Also, the less information is retained during the analysis, the larger the overapproximation becomes, potentially leading to false positive results. Currently, selecting an appropriate abstraction is left to the user, requiring considerable expertise and manual effort.

In this paper, we  address this problem by introducing an automated refinement approach that dynamically adapts abstraction levels during symbolic-execution-based IFA using heuristic guidance. Our approach is twofold: First, we propose to reduce the symbolic state space that is explored during the symbolic execution using \emph{exploration refinement}. Second, we propose a dedicated \emph{information flow refinement} to eliminate spurious information flow caused by overapproximation whenever illegal information flow is detected. The approach is based on the second author's Master's thesis \cite{Ernst2026}.
The underlying key idea of the exploration refinement is to track additional statements and variables whenever an unknown branching decision leads to a \enquote{large} diversion in the state space. To decide which additional statements and variables are tracked, we propose two novel heuristics: The \emph{partial descendants heuristic} counts the number of successor states of a given branching node that lie on one branch, but not on all of them. The \emph{splitters heuristic}  identifies branching conditions that cause diverging scheduling states if left untracked.
For information flow refinement, on the other hand, we propose a novel heuristic called \emph{path resolution}, which tries to specifically refine nodes on a given path where potential illegal information flow is detected to exclude spurious 
information flow during IFA. 
 Overall, the proposed heuristics reduce the need for expert-defined abstractions while improving both the efficiency and accuracy of symbolic-execution-based IFA.

We have implemented our approach for the system design language SystemC \cite{SystemC2023}, which is an industrially widely used language for the modeling and simulation of hardware/software co-designs and shared bus architectures  \cite{ghenassia2006transaction}. We demonstrate the feasibility of our approach with several case studies, which show that the automatically chosen abstractions enable precise and efficient IFA without requiring manually defined abstractions. 
The implementation and a full reproduction package are available as an artifact to this paper \cite{artifact}.

The rest of this paper is structured as follows: In \Cref{sec:background} we introduce IFA, SystemC, and the approach from \cite{Becker-Kupczok2024a}. We then introduce our novel abstraction refinement heuristics in \Cref{sec:main}, before presenting the experimental evaluation in \Cref{sec:evaluation}. 
We discuss related work in \Cref{sec:related_work} and conclude in \Cref{sec:conclusion}.

\section{Background}\label{sec:background}

This section introduces the necessary background for the remainder of this paper, namely IFA, SystemC, and the approach from \cite{Becker-Kupczok2024a}, on which this work builds. 

\subsection{Information Flow Analysis}\label{subsec:background_informationFlow}

The goal of IFA is to identify all program locations (particularly all inputs) that can influence the execution of a given program location (usually an output). This information can be used to analyze properties such as non-interference~\cite{Goguen1982}, which can be used to establish confidentiality or integrity.

IFA can be carried out using security type systems \cite{Volpano1996} or program dependence graphs (PDGs) \cite{Hsieh1992}. Both techniques generally have the same precision \cite{Mantel2013}. A PDG contains a node for every program location in a method and edges for all possible influences between program locations. These edges are typically induced by data and control dependencies. 
The PDG can then be used to compute the backwards slice \cite{Weiser1984} of any output, i.e. the set of all nodes from which there exists a path in the PDG to the output node (the slicing criterion).

\subsection{SystemC}

SystemC is a system-level design language and a framework for HW/SW co-simulation. Its semantics are informally defined in an IEEE standard \cite{SystemC2023}. It is implemented as a C++ class library, providing language elements for the description of hardware and software at various levels of abstraction.
It also features an event-driven simulation kernel.
A SystemC design consists of a set of communicating processes, triggered by events or timing delays and interacting through channels. SystemC is especially effective for modeling shared bus architectures because it can accurately simulate bus communication, arbitration, contention, and timing to evaluate system performance before hardware implementation.

Like typical hardware simulators, SystemC models concurrency by imposing a partial order on parallel processes using delta-cycles: During simulation, the scheduler operates in phases. In the evaluation phase, a ready process is selected non-deterministically and executed until it terminates or reaches a wait-statement. This is repeated until no more processes are ready, in which case the scheduler switches to the update/notify phase. There, values written to signals or channels during the previous evaluation phase become visible to other processes. At the end of the update/notify phase, a delta delay passes. This is a special unit of simulation time which is treated as greater than zero but smaller than any real delay, and each elapsed delta delay constitutes one delta cycle. Afterwards, if no process is ready, simulation time advances to the earliest point at which any process becomes ready. Finally, a new evaluation phase is started.

SystemC uses a cooperative scheduler, which means that threads only yield control voluntarily and are executed atomically between waits.

\subsection{Symbolic Execution for IFA of Embedded Systems}

Symbolic execution executes a program with symbolic values \cite{King1976}, allowing many executions with different inputs or non-deterministic choices to be simulated in one or few passes.
Classical symbolic execution represents symbolic values as terms and expressions derived from the program. 
\cite{Becker-Kupczok2024a} uses a more general definition where symbolic values range from concrete values, over elements of abstract domains \cite{Cousot1977} (e.g., intervals), to completely unknown values.

\begin{figure}[tbp]
    \centering
    \resizebox{\linewidth}{!}{\input{img/approach_24}}
    \caption{Symbolic Execution for IFA of Embedded Systems
    \cite{Becker-Kupczok2024a}.}
    \label{fig:approach_24}
\end{figure}
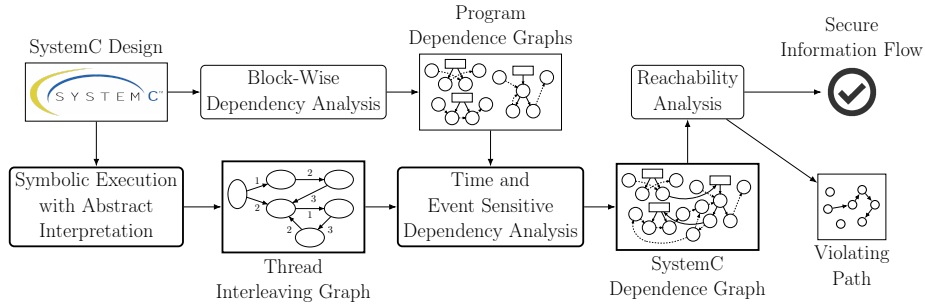

The approach from \cite{Becker-Kupczok2024a} (visualized in \Cref{fig:approach_24}) exploits the fact that code in SystemC 
forms atomic blocks. It creates PDGs for these blocks using a block-wise depencency analysis, and uses a symbolic execution with abstract interpretation to find an overapproximation of all their possible interleavings captured as a thread interleaving graph. It uses this information to combine all the PDGs into a \emph{SystemC dependence graph} (SysCDG), which captures all data, timing, event, and other control dependencies between blocks. IFA can then be performed by
reachability analysis on the SysCDG. The analysis either finds a violating path where illegal information flow might occur or concludes that the system is secure.

\section{Heuristic Abstraction Refinement}\label{sec:main}

\begin{figure}[tbp]
    \centering
    \resizebox{\linewidth}{!}{\input{img/approach}}
    \caption{Precise IFA with Automated Abstraction Refinement.} 
    \label{fig:approach}
\end{figure}
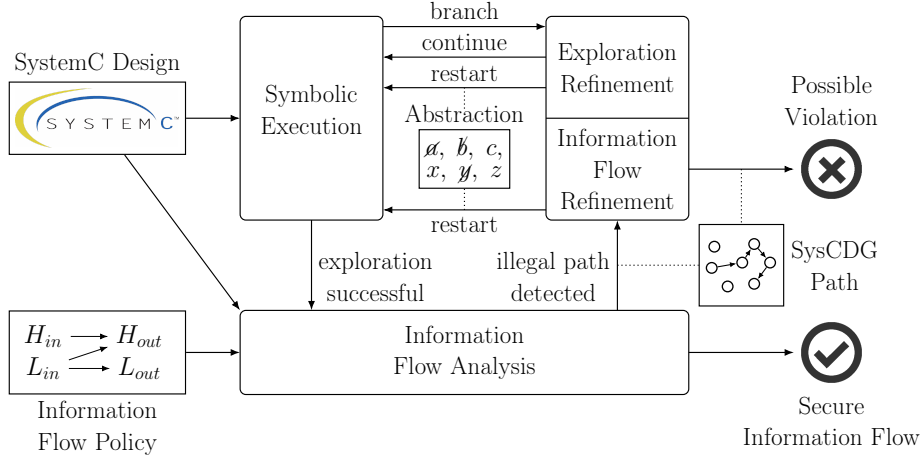

The symbolic execution used for precise IFA in \cite{Becker-Kupczok2024a} strongly relies on an appropriate user-defined abstraction -- \ie the user needs to decide which variables are kept concrete and which are abstracted. This choice has a significant impact on the analysis. On the one hand, it affects the runtime (and de-facto termination) of the symbolic execution. On the other hand, too much abstraction  reduces the precision of the IFA. Consequently, selecting a suitable abstraction requires considerable expertise and manual effort.

To address this problem, we propose heuristically selecting abstraction levels based on dependencies between states and detected potential information leakage. During symbolic execution, we detect whenever the state space branches and identify refinements that can avoid unnecessary branching. Since these refinements are applied during symbolic exploration, we refer to this process as \emph{exploration refinement}. To determine suitable refinement candidates, we propose two heuristics:
\begin{enumerate} 
\item \emph{Partial descendants}, which counts abstract states reachable from some, but not all, branches originating from the current state.
\item \emph{Splitters}, which identifies control conditions that cause splits in the scheduler state space if left untracked.
\end{enumerate}

The goal of exploration refinement is to reduce the symbolic state space explored during symbolic execution, which is subsequently used to construct the thread interleaving graph and the SystemC dependence graph. A smaller state space accelerates symbolic execution and simplifies dependence analysis.

Besides the scalability issues that result from large symbolic state spaces, a further problem of IFA for timed concurrent systems using symbolic execution together with abstract interpretation is that due to the overapproximation, it might not be able to conclude the absence of illegal information flow due to spurious paths in the SysCDG. To counteract this problem, we propose \emph{information flow refinement}, which aims at eliminating spurious information flow caused by over-
approximation whenever illegal information flow is detected.
For \emph{information flow refinement}, we propose a \emph{path resolution heuristics}, which tries to specifically refine nodes on a given path where potential illegal information flow is detected, for example by
identifying control conditions on this path
and choosing these conditions as candidates for refinement. 

Our overall approach to automate the process of finding a suitable abstraction using exploration and information flow refinement is visualized in \Cref{fig:approach}.
Initially, we perform a \emph{Symbolic Execution} on a given \emph{SystemC Design} with a trivial abstraction, e.g. an abstraction where all values are symbolically captured, i.e. assumed to be completely unknown. 
Whenever we detect a \emph{branch} during the \emph{Symbolic Execution}, we perform \emph{Exploration Refinement} on the current \emph{Abstraction}. For \emph{Exploration Refinement}, we employ 
the \emph{partial descendants heuristic}, which refines the abstraction if the branch leads to a \enquote{large} number of different followup states, and the \emph{splitters heuristic}, which refines the abstraction if the branch leads to a difference in the scheduling state. If one of the heuristics indicates that a refinement should be applied, we \emph{restart} the \emph{Symbolic Execution}. Otherwise, we just  \emph{continue} with the previously used \emph{Abstraction}.

Once the \emph{Symbolic Execution} has successfully completed with a given abstraction (\emph{exploration successful}),
we perform \emph{Information Flow Analysis} for a given \emph{Information Flow Policy} that specifies which high (i.e. confidential or untrusted)
inputs $H_{in}$ and low (i.e. public or trusted) inputs $L_{in}$ may influence which high outputs $H_{out}$ and low outputs $L_{out}$. 
If a violating path is detected (\emph{illegal path detected}), we employ the \emph{path resolution heuristic} to attempt \emph{Information Flow Refinement} with the goal of finding an \emph{Abstraction} where this violating path does not occur, meaning that the illegal information flow is not actually possible. This triggers a new symbolic execution (\emph{restart}), possibly leading to more exploration refinement in between information flow refinements. The analysis terminates once either no violating paths remain or no further promising refinement candidates can be identified.

Our approach is compatible with different kinds of abstractions, some of which are discussed in \Cref{subsec:kinds_of_abstractions}. Then, we introduce exploration and information flow refinement in \Cref{subsec:exploration_refinement,subsec:information_flow_refinement}.

\subsection{Kinds of Abstractions}\label{subsec:kinds_of_abstractions}

In principle, any abstraction in the sense of abstract interpretation \cite{Cousot1977} can be used in our approach. However, all  heuristics presented in this paper rely on tracking or not tracking certain \emph{values}. In the simplest form, these values are variable values, such that an abstraction is defined by a set of variables whose values are tracked.
This approach may, however, be unnecessarily coarse. 
A variable may initially be worth tracking, e.g. because it appears in a branching condition affecting potential information flow, but later assume many different values that are no longer relevant. In such cases, it is preferable to track the variable only until the branching decision has been made.

Therefore, our abstractions are typically characterized by a set of \emph{variable assignments} whose values are tracked resp. not tracked. Conceptually, this corresponds to viewing the program in static single assignment form \cite{Cytron1986}, although no explicit transformation is performed.

\subsection{Exploration Refinement using Partial Descendants and Splitters}\label{subsec:exploration_refinement}

The goal of the exploration refinement is to identify an abstraction that reduces the size of the symbolically explored  state space, allowing symbolic execution to terminate successfully in acceptable time. In general, higher abstractions yield smaller state spaces than lower (more concrete) ones. Therefore, we begin with the highest possible abstraction, tracking no values except constants. We then incrementally refine that abstraction whenever doing so is expected to reduce rather than increase the size of the explored state space.

This is potentially the case whenever the execution reaches a branching point where it cannot decide which branch to take. Without refinement, all branches must be explored separately. If they merge again quickly, this incurs little overhead and refinement would be unnecessary. However, if they don't, refining the abstraction to track the values determining the branching decision substantially reduce the exploration effort.
These two cases are very difficult to distinguish without already knowing the state transition graph (the \emph{thread interleaving graph} from \cite{Becker-Kupczok2024a}), which the symbolic execution aims to construct. Therefore, we tackle the problem heuristically.
The two heuristics we propose in this paper are the \emph{partial descendants} and \emph{splitters} heuristics, introduced in the following sections. 

Note that whenever we track a value, we transitively extend that tracking to all values used to compute it, because assigning an unknown value to a tracked variable would make it unknown again, defeating the point of tracking it.

\subsubsection{Partial Descendants Heuristic.}\label{subsubsec:partial_descendants}

If the complete state transition graph were available, the usefulness of refining a branch decision could be assessed directly by determining the number of states that are reachable from only a subset of the outgoing branches of a branching state. This information could be used to decide whether tracking the variables involved in the branch condition is likely to reduce the explored state space. Since the complete state transition graph is not available during symbolic exploration, such an exact analysis is infeasible. Instead, we approximate this quantity using the partially explored state transition graph and estimate whether the number of such states exceeds a predefined threshold. This approximation forms the basis of the \emph{partial descendants heuristic}.

Intuitively, a state is a \emph{partial descendant} (PD) of a branching state, if it is reachable from at least one of the branches without returning to the branching state, but not from all of them. Formally, we define this notion as follows.
\begin{definition}[Partial Descendants]
    Let $G = (V, E)$ be a directed graph with $V$ a set of abstract states and $E \subseteq V \times V$ a set of transitions between them. Let $P = \set{(v_1, \ldots, v_n) \mid \parens{v_1, v_2}, \ldots, \parens{v_{n-1}, v_n} \in E}$
    be the set of finite paths in $G$. For $v, v' \in V$, we call $v'$ a \emph{partial descendant of $v$} iff there exist distinct $v_1, v_2 \in V$ with $\parens{v, v_1}, \parens{v, v_2} \in E$ such that there is a path $\parens{v_1, \ldots, v'} \in P$ not containing $v$, but there is no path $\parens{v_2, \ldots, v'} \in P$ not containing $v$.
\end{definition}
This definition is applicable to both fully explored and partially constructed state transition graph.
The partial descendants heuristic monitors the number of PDs of every state during the exploration. If this number exceeds a predefined threshold, the abstraction is refined to track the values controlling the branch (\eg those occuring in an if-condition).
If a branching statement is evaluated in multiple abstract states, we use the maximum number of PDs among these states rather than their sum. If a condition is evaluated in many states that have only a few PDs each, 
resolving the branch can reduce the state space only marginally, whereas the additional tracking may substantially increase it. 

Since the heuristic operates on the partially explored graph, it can in some cases overestimate the number of PDs. This occurs when a path from at least one branch to a node is already explored, but the paths from the remaining branches to that same node are yet to be discovered. 
To mitigate this effect, we perform symbolic execution using breadth-first search. Although this does not eliminate the issue, the remaining cases are unlikely to be problematic: either the missing path is significantly longer, making refinement potentially beneficial anyway, or the actual number of PDs is only slightly below the threshold, in which case refining the branch remains reasonable.

\subsubsection{Splitters Heuristic.}\label{subsubsec:splitters}

Similar to the partial descendants heuristic, the splitters heuristic identifies control conditions that cause a large split in the state space if left untracked. However, instead of estimating this effect explicitly, it relies on a simpler criterion:
Since the scheduler state, i.e. which processes are waiting for which events or timing delays, is typically the most relevant part of the state for identifying possible interleavings of atomic blocks, the splitters heuristic tracks a control condition only if leaving it unknown changes the scheduler state at the end of an atomic block. This idea is inspired by the refinement heuristic of \cite{Tasche2024a}.

Although less precise than the PD heuristic, this approach is considerably cheaper, while still working well for typical SystemC designs. Because the heuristic is evaluated locally at the end of each atomic block, it requires no additional state and can be applied on the fly without being affected by subsequent states.

As an exception, we propose \emph{not} to track the concrete values of primitive channels in SystemC in certain cases. Some such channels, such as \texttt{sc\_signal}, notify an event only when written with a value different from their current one. This causes a split in the scheduler state based on the channel's value, but tracking these concrete values typically increases the state space much more than it reduces it. However, if a channel value is subsequently used in another control condition that also causes a split, then our heuristic would decide to track it.

\subsection{Information Flow Refinement}\label{subsec:information_flow_refinement}

\begin{algorithm}[tb]
\caption{Cumulative Information Flow Refinement}\label{alg:if_refinement}
\begin{algorithmic}[1]
    \Require SystemC design $D$, initial abstraction $A_0$ (set of tracked values, can be $\emptyset$)
    \Ensure potentially violating path in SysCDG, or $\bot$ if none exists
    \State $A \gets \texttt{exploration\_refinement}(D, A_0)$ \MComment{use exploration refinement to find first abstraction for the symbolic execution}
    \State $T \gets \texttt{duration\_of\_execution}()$ \Comment{take the time for the successful execution}
    \State $S \gets \texttt{construct\_syscdg}(D, A)$ \Comment{construct SysCDG with the found abstraction}
    \While{$\texttt{violating\_paths}(S) \neq \emptyset$} \Comment{repeat while violating path exists}
        \State \textbf{choose} $P \in \texttt{violating\_paths}$(S) \Comment{choose any violating path}
        \For{$R \in \texttt{refinement\_candidates}(S, P)$} \Comment{iterate over refinement candidates}
            \State $A' \gets \texttt{exploration\_refinement}(D, A \union R, \texttt{timeout}(T))$
            \MComment{try to find a new abstraction for the symbolic execution based on the refined one, using a relative timeout}
            \State $T' \gets \texttt{duration\_of\_execution}()$ \Comment{take the time again}
            \If{$A' = \bot$} \Comment{if the exploration was unsuccessful, reject the candidate}
                \State \textbf{Continue for loop}
            \EndIf
            \State $S' \gets \texttt{construct\_syscdg}(D, A')$ \Comment{construct a new SysCDG}
            \If{$\lnot \texttt{is\_path\_present}(S', P)$}
                \MComment{if the candidate eliminates the path, then update abstraction, time, and SysCDG, and continue with the next path (if any)}
                \State $A \gets A', T \gets \max\parens{T, T'}, S \gets S'$
                \State \textbf{Continue while loop}
            \EndIf
        \EndFor
        \State \Return $P$ \Comment{if out of candidates, return possibly violating path}
    \EndWhile
    \State \Return $\bot$
\end{algorithmic}
\end{algorithm}

The goal of information flow refinement is to eliminate spurious information flow whenever a potential security violation is detected. 
To this end, we iteratively apply the path resolution heuristic to remove paths from the SysCDG that induce illegal information flow until either no violating path is left (and we report secure information flow), or no promising refinement candidates are left.

Refinements identified by the path resolution heuristic for different paths can be combined in two ways: First, they can be accumulated, meaning that the final abstraction may resolve all violating paths simultaneously. Second, each path can be resolved individually. We refer to these variants as \emph{cumulative} and the \emph{non-cumulative information flow refinement}, respectively.

The cumulative procedure is shown in \Cref{alg:if_refinement}. To prevent an unfavorable refinement candidate from causing a state space explosion and excessive symbolic execution times, we measure the duration of the first successful symbolic execution and use this time as a basis for a timeout on any subsequent execution. If the timeout is exceeded, the candidate is rejected, whereas successful executions update the reference runtime. We propose to set the timeout to $a \cdot T + b$ (for a base duration $T$), and use $a = 5$ and $b = 10s$ as default values.

The non-cumulative procedure is similar, but differs in two aspects: it
maintains a set $\mathcal{S}$ of SysCDGs instead of a single SysCDG $S$, and
lines~5--6 consider only violating paths present in \emph{all} graphs, as
clarified below. When a path $P$ is resolved, \ie an abstraction is found whose
SysCDG no longer contains $P$, that SysCDG is added to $\mathcal{S}$. Unlike in
the cumulative variant, neither the abstraction $A$ nor the base timeout
$T$ is updated. Since $\mathcal{S}$ then contains a graph without $P$, this path
cannot be selected again, even if the abstraction is reset. This can
keep the abstraction smaller, reducing runtime or making the analysis feasible.

\Cref{alg:if_refinement} and our description of the non-cumulative procedure leave two things unspecified: The selection of refinement candidates, and the determination of whether a path is present in a different or even in multiple SysCDGs (which can have different sets of nodes). We discuss these in the following sections.

Note that, in order for the information flow refinement to always terminate, the exploration refinement must be monotone, \ie never remove values from being tracked. Both our exploration heuristics satisfy this requirement.

\subsubsection{Path Resolution Heuristic.}

For selecting refinement candidates, we propose the path resolution heuristic. The idea is that removing any node from a path resolves that path. This means that at least the value of any condition controlling a node on the path is a candidate for refinement.

Since, when selecting candidates for the information flow refinement, we already have a SysCDG, these controlling nodes are easy to find by simply traversing incoming control dependency edges from the node in question, as well as transitively from already identified controlling nodes. Apart from classical controlling nodes like if-statements, we also treat wait-statements and event notifications as controlling the atomic blocks triggered by them.

\begin{figure}[tb]
    \begin{subfigure}[b]{0.475\textwidth}
        \centering
        \input{img/path_resolution_sender}
        \caption{Process writing low and high information to \texttt{val}, setting \texttt{flag} accordingly.}
        \label{fig:examples_path_resolution_sender}
    \end{subfigure}\hfill
    \begin{subfigure}[b]{0.475\textwidth}
        \centering
        \input{img/resolution_data_of_control}
        \caption{Path that can be resolved by tracking data dependencies of a controlling node.}
        \label{fig:examples_path_resolution_data_of_control}
    \end{subfigure}\\
    \begin{subfigure}[b]{0.475\textwidth}
        \centering
        \input{img/resolution_slice_of_control}
        \caption{Path that can be resolved by tracking all dependencies of a controlling node.}
        \label{fig:examples_path_resolution_slice_of_control}
    \end{subfigure}\hfill
    \begin{subfigure}[b]{0.475\textwidth}
        \centering
        \input{img/resolution_slice_of_slice}
        \caption{Path that can be resolved by tracking all dependencies of a dependency node.}
        \label{fig:examples_path_resolution_slice_of_slice}
    \end{subfigure}
    \begin{tikzpicture}[remember picture,overlay,codenode/.style n args={1}{draw=#1,fill=#1,fill opacity=0.1,thick,rounded corners,minimum height=0.375cm},mypath/.style n args={1}{draw=#1,fill=#1}]
        \node[codenode=magenta,minimum width=2.85cm] (high_write) at ([yshift=0.065cm,xshift=0.175cm]pic cs:high_write) {};
        \node[codenode=magenta,minimum width=2.85cm] (low_read_b) at ([yshift=0.085cm,xshift=0.0cm]pic cs:low_read_b) {};
        \node[codenode=magenta,minimum width=2.85cm] (low_read_c) at ([yshift=0.065cm,xshift=0.0cm]pic cs:low_read_c) {};
        \node[codenode=magenta,minimum width=3.1cm] (high_read_d) at ([yshift=0.085cm,xshift=0.085cm]pic cs:high_read_d) {};
        \node[codenode=magenta,minimum width=3.1cm] (low_write_d) at ([yshift=0.085cm,xshift=-0.11cm]pic cs:low_write_d) {};
        
        \node[codenode=cyan,minimum width=2.85cm] (flag_check_b) at ([yshift=0.085cm,xshift=-0.18cm]pic cs:flag_check_b) {};
        \node[codenode=cyan,minimum width=2.7cm] (wait_c) at ([yshift=0.085cm,xshift=-0.1cm]pic cs:wait_c) {};
        \node[codenode=cyan,minimum width=2.85cm] (flag_check_d) at ([yshift=0.085cm,xshift=-0.0cm]pic cs:flag_check_d) {};
        \node[codenode=cyan,dotted,minimum width=1.85cm] (overwrite_d) at ([yshift=0.085cm,xshift=-0.09cm]pic cs:overwrite_d) {};
        
        \node[codenode=darkgreen,minimum width=3.25cm] at ([yshift=0.085cm,xshift=0.175cm]pic cs:flag_set) (flag_set) {};
        \node[codenode=darkgreen,dashed,minimum width=2.9cm] at ([yshift=0.075cm,xshift=0.185cm]pic cs:flag_check) (flag_check) {};
        \node[codenode=darkgreen,dashed,minimum width=4.9cm] at ([yshift=0.085cm,xshift=0.175cm]pic cs:notify) (notify) {};

        \path[mypath=magenta,thick] (high_write) edge[-Latex] (low_read_b);
        \node[draw=none,magenta,rotate=90] (stop_c_1) at ([xshift=1cm,yshift=-1.2cm]high_write.-5) {$\cdots$};
        \node[draw=none,magenta,rotate=85] (stop_c_2) at ([xshift=1cm,yshift=1.2cm]low_read_c.5) {$\cdots$};
        \path[mypath=magenta,thick] ([xshift=-0.0625cm,yshift=0.0625cm]high_write.south east) edge[bend left] (stop_c_1);
        \path[mypath=magenta,thick] (stop_c_2) edge[-Latex,bend left] ([xshift=-0.045cm,yshift=-0.045cm]low_read_c.north east);
        \node[draw=none,magenta,rotate=110] (stop_d_1) at ([xshift=1.5cm,yshift=-1cm]high_write.-5) {$\cdots$};
        \node[draw=none,magenta,rotate=100] (stop_d_2) at ([xshift=-0.75cm,yshift=1cm]high_read_d.175) {$\cdots$};
        \path[mypath=magenta,thick] (high_write.east) edge[bend left] (stop_d_1);
        \path[mypath=magenta,thick] (stop_d_2) edge[-Latex,bend right] ([xshift=0.045cm,yshift=-0.045cm]high_read_d.north west);
        \draw[magenta,thick] plot[smooth,tension=1] coordinates { (high_read_d.west) ([xshift=-0.25cm]flag_check_d.west) ([xshift=-0.03cm,yshift=0.027cm]low_write_d.west)};
        \path[mypath=magenta,thick] ([xshift=-0.001cm,yshift=0.0009cm]low_write_d.west) edge[-Latex] (low_write_d.west);
        
        \draw[cyan,thick] plot[smooth,tension=1] coordinates { ([xshift=-0.05cm]flag_check_b.5) ([xshift=-0.0625cm,yshift=-0.0625cm]flag_check_b.5 -| low_read_b.5) ([yshift=0.15cm]low_read_b.5)};
        \path[mypath=cyan,thick] ([xshift=-0.00001cm,yshift=0.001cm]low_read_b.5) edge[-Latex] ([yshift=-0.025cm]low_read_b.5);
        \draw[cyan,thick] plot[smooth,tension=1] coordinates { (wait_c.west) ([xshift=-0.25cm]low_read_c.north-|low_read_c.west) (low_read_c.west)};
        \path[mypath=cyan,thick] ([xshift=-0.001cm,yshift=0.0006cm]low_read_c.west) edge[-Latex] (low_read_c.west);
        \path[mypath=cyan,thick] (flag_check_d) edge[-Latex] (overwrite_d);
        \path[mypath=cyan,thick] (overwrite_d.south) edge[-Latex,bend left] ([xshift=-0.0625cm,yshift=0.0625cm]low_write_d.south east);
        
        \path[mypath=darkgreen,thick] (flag_set.east) edge[-Latex] (flag_check_b.west);
        \draw[darkgreen,thick,dashed] plot[smooth,tension=1] coordinates { ([xshift=0.0625cm,yshift=-0.0625cm]flag_set.north west) ([xshift=-0.25cm,yshift=0.125cm]flag_check.north-|flag_check.west) (flag_check.west)};
        \path[mypath=darkgreen,thick] ([xshift=-0.001cm,yshift=0.0008cm]flag_check.west) edge[-Latex] (flag_check.west);
        \path[mypath=darkgreen,thick,dashed] ([xshift=0.0625cm,yshift=0.0625cm]flag_check.south west) edge[-Latex,bend right] ([xshift=0.0625cm,yshift=-0.0625cm]notify.north west);
        \node[draw=none,darkgreen,rotate=110] (stop_c_3) at ([xshift=-0.25cm,yshift=-2.25cm]notify.south west) {$\cdots$};
        \node[draw=none,darkgreen,rotate=140] (stop_c_4) at ([xshift=-0.5cm,yshift=1cm]wait_c.north east) {$\cdots$};
        \path[mypath=darkgreen,thick,dashed] ([xshift=0.0625cm,yshift=0.0625cm]notify.south west) edge[bend right] (stop_c_3);
        \path[mypath=darkgreen,thick,dashed] (stop_c_4) edge[-Latex,bend left] ([xshift=-0.0625cm,yshift=-0.0625cm]wait_c.north east);
        \node[draw=none,darkgreen,rotate=110] (stop_d_3) at ([xshift=1.25cm,yshift=-1.1cm]high_write.-5) {$\cdots$};
        \node[draw=none,darkgreen,rotate=100] (stop_d_4) at ([xshift=-1.0cm,yshift=0.75cm]flag_check_d.north west) {$\cdots$};
        \path[mypath=darkgreen,thick] ([xshift=-0.0625cm,yshift=0.0625cm]flag_set.south east) edge (stop_d_3);
        \path[mypath=darkgreen,thick] (stop_d_4) edge[-Latex,bend right] ([xshift=-0.0625cm,yshift=-0.0625cm]flag_check_d.west);
    \end{tikzpicture}
    \caption{Examples of paths that require different resolution techniques. \Cref{fig:examples_path_resolution_sender} shows a process writing (potentially high) information to the variable \texttt{val}, the other figures show processes reading that variable and potentially writing to a low output. The paths are shown in magenta, the nodes used to resolve them in blue, and their tracked influences in green. The dashed green nodes only belong to the influences of \ref{fig:examples_path_resolution_slice_of_control}.}
    \label{fig:examples_path_resolution}
\end{figure}
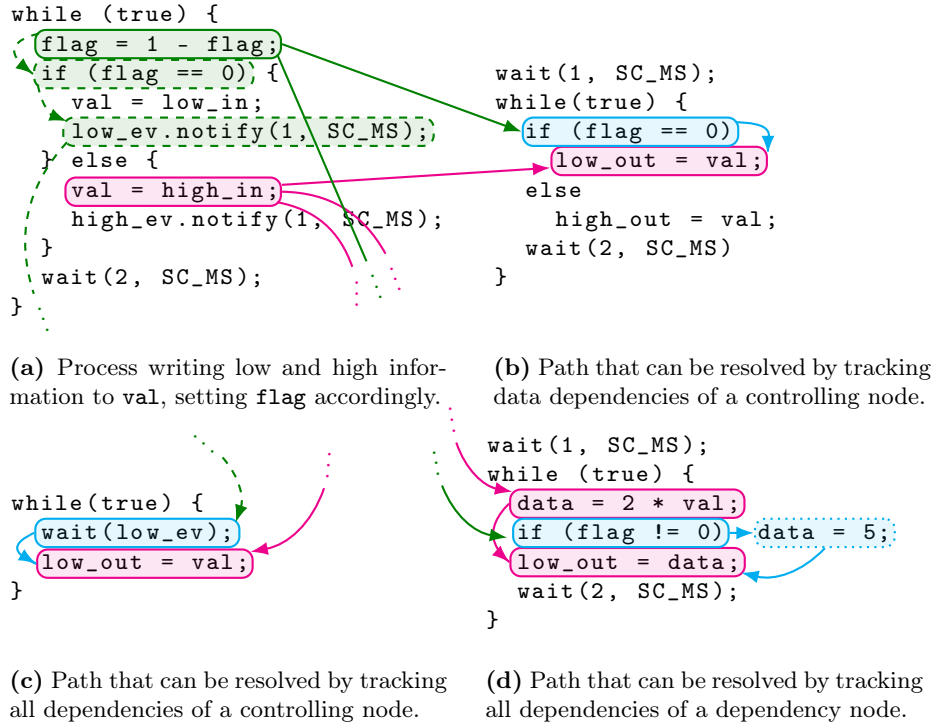

For a given controlling node that may resolve the path, it may be sufficient to ensure that the value of its condition (or, in the case of wait-statements and event notifications, its parameters) is known whenever it is encountered. This can be achieved by tracking the values of all its transitive data dependencies -- \ie by taking the backwards slice of the controlling node, but limited to only traversing data dependency edges, and tracking all assignments made in nodes in this slice. Such a case is illustrated in \Cref{fig:examples_path_resolution_data_of_control}: \texttt{low\_out = val} is part of the potentially violating path. It is control dependent on \texttt{if (flag == 0)}, which is data dependent on \texttt{flag = 1 - flag}. Tracking \texttt{flag} will resolve the path.

In some cases, considering only data dependencies of controlling nodes is not sufficient, but considering all their dependencies is. Such a case is illustrated in \Cref{fig:examples_path_resolution_slice_of_control}: \texttt{low\_out = val} is again part of the path, and it is control dependent on \texttt{wait(low\_event)}. However, this has no data dependencies worth tracking. Instead, it has a control dependency on \texttt{low\_ev.notify(1, SC\_MS)}, which in turn is control dependent on \texttt{if (flag == 0)}. This does have a data dependency that, if trackked, resolves the path (\texttt{flag = 1 - flag}). Since, by definition, this approach results in a superset of values tracked compared to only considering data dependencies of controlling nodes, we prefer the latter whenever it suffices.

Finally, in relatively rare cases, only considering controlling nodes of nodes on the path can be insufficient to resolve it. This may be the case if, \eg, a data dependency is overwritten with a low value, but the current abstraction is insufficient to show this to be the case. Such a situation is illustrated in \Cref{fig:examples_path_resolution_slice_of_slice}: \texttt{low\_out = data} is part of the path, and has no non-trivial control dependencies. However, it does have a data dependency on \texttt{data = 5}, which is control dependent on \texttt{if (flag != 0)}. Tracking all dependencies of this node (which includes \texttt{flag = 1 - flag}) resolves the path, because the perceived data dependency between \texttt{data = 2 * val} and \texttt{low\_out = data} is killed whenever \texttt{val} is high. Because this is relatively rare, we first consider the other two possibilities.

\begin{algorithm}[tb]
\caption{Path Resolution Heuristic}\label{alg:path_resolution}
\begin{algorithmic}[1]
    \Require SysCDG $S$, path $P$
    \Ensure List of refinement candidates (sets of tracked values)
    \State $L \gets \text{\Call{select}{$S$, $P$, \texttt{controlling\_slice}, \texttt{data\_slice}}}$ \MComment{select data dependencies of controlling nodes}
    \State $L \gets L \circ \text{\Call{select}{$S$, $P$, \texttt{controlling\_slice}, \texttt{backwards\_slice}}}$
    \MComment{select other dependencies of controlling nodes}
    \State $L \gets L \circ \text{\Call{select}{$S$, $P$, \texttt{backwards\_slice}, \texttt{backwards\_slice}}}$ \MComment{select any dependencies of any nodes}
    \Statex
    \Function{select}{$S$, $P$, \texttt{get\_candidates}, \texttt{get\_influences}}
        \State $L \gets \texttt{[]}$ \Comment{initialize result with empty list of candidates}
        \For{$N \in P$} \Comment{iterate through the path}
            \For{$C \in \text{\Call{get\_candidates}{$S$, $C$}}$} \Comment{iterate over candidates to resolve $N$}
                \State $R \gets \emptyset$ \Comment{initialize refinement candidate with empty set of values}
                \For{$I \in \text{\Call{get\_influences}{$S$, $C$}}$} \Comment{iterate over nodes influencing $C$}
                    \State $R \gets R \union \texttt{written\_values}(I)$ \Comment{add values written in $I$ to tracking set}
                \EndFor
                \State $L \gets L \circ R$ \Comment{add candidate set to result list}
            \EndFor
        \EndFor
        \State \Return $L$ \Comment{return result list}
    \EndFunction
\end{algorithmic}
\end{algorithm}

Taken all together, our proposed path resolution heuristic for information flow refinement is shown in \Cref{alg:path_resolution}. The functions \texttt{controlling\_slice}, \texttt{data\_slice}, and \texttt{backwards\_slice} return the backwards slice of the given node in the graph, but traversing only control dependency edges, data dependency edges, or all edges, respectively.

Concerning the order in which nodes are traversed (and, therefore, in which refinement candidates are produced), we propose to traverse the path from its source to its target, because nodes earlier in the path tend to have smaller backwards slices (leading to smaller refinement candidates). For traversing the candidates slices, we propose a breadth first search starting right after the node itself, because usually, control conditions closer to the path are more likely to resolve it. Traversal order for the influence slices is irrelevant, because all their values are combined in a single refinement candidate.

Note that the refinement candidates don't have to be computed en bloc, but can also be determined on demand, whenever a new candidate is required.

\subsubsection{Identifying Paths in Different SysCDGs.}

The SysCDG as constructed according to \cite{Becker-Kupczok2024a} consists of PDGs corresponding to executions of atomic blocks, and connected according to to their possible interleavings. Changes to the abstraction affect these possible interleavings (that is the purpose of the refinement). Therefore, after a refinement, the SysCDG may look drastically different than before. That makes it non-trivial to test whether a potentailly violating path from one SysCDG still exists in the new SysCDG (\ie the implementation of \texttt{is\_path\_present} from line 14 in \Cref{alg:if_refinement}), or to find a single path that appears in multiple SysCDGs.

One could assume that, since our refinement is monotonic, the set of possible interleavings only ever decreases, and that the new SysCDG therefore must be a subgraph of the old. However, this is not the case: Even though the number of actually possible interleavings given the abstraction always decreases, a set of interleavings that may previously have been captured with just one sequence of abstract states may subsequently require several such sequences, because each abstract state now covers a smaller set of concrete states. The new SysCDG may therefore be neither a sub- nor a supergraph of the old one.

We solve this problem by mapping paths to \emph{path signatures}. A path signature is a sequence of \emph{node signatures}, where a node signature serves to identify a node across SysCDGs. A straight forward example for concrete node signatures (and the one we use for our implementation) is to map nodes to the program locations they represent. If a path passing the same program locations in the same order exists in another graph (even if the abstract state transitions the nodes belong to look different), then it is likely the same path. Conversely, if such a path doesn't exist, \eg because some location which is part of the path is not executed with the abstraction inducing the other graph, then there likely isn't actually any information flow following along that path: If there was, \ie if a concrete execution of the system existed where this information flow took place, then any sound abstraction would also induce such a path.

With this, the implementation of \texttt{is\_path\_present} becomes trivial: Perform a straight forward path finding on the new graph considering only successor nodes that have the same signature as the next node on the path.

In contrast, finding a common violating path across SysCDGs is a bit more complex. We do this by taking a subgraph of the product graph of all considered SysCDGs, where only nodes with the same signatures may appear in one product node. We formalize this as follows.
\begin{definition}[Signature Product Graph]
    Let $\parens{G_i}_{i \in \set{1, \ldots, n}}$ with $G_i = \parens{V_i, E_i}$ be a sequence of SysCDGs, let $V = \bigUnion_{i \in \set{1, \ldots, n}} V_i$, and let $\sign: V \to S$ map each node to a signature from a set $S$. Then the \emph{signature product graph of $\parens{G_i}_{i \in \set{1, \ldots, n}}$} is the graph $\hat{G} = \parens{\hat{V}, \hat{E}}$ with:
    \begin{align*}
        \hat{V} &= \set{\parens{v_1, \ldots, v_n} \in V_1 \times \ldots \times V_n \with \signOf{v_1} = \ldots = \signOf{v_n}} \ \text{and} \\
        \hat{E} &= \set{\parens{\parens{v_1, \ldots, v_n}, \parens{v'_1, \ldots, v'_n}} \in \hat{V} \times \hat{V} \with \parens{v_1, v'_1} \in E_1 \land \ldots \land \parens{v_n, v'_n} \in E_n}
    \end{align*}
\end{definition}
A common path in a set of SysCDGs is a path in their signature product graph.

Looking for common paths is more precise than just checking for the presence of an old path in a new graph: Consider the case of a path where information flows into a loop, from one iteration to the next, and then out after termination, where the iterations are separated by wait statements (\ie are forming individual atomic blocks). If the amount of iterations is unknown when the first SysCDG is created, then the path found will contain \eg one iteration. If, after refinement, the loop is known to be executed exactly four times, then that path is no longer present in the new SysCDG. However, there still is a commmon path in both SysCDGs, because the path with four iterations is also present in the old SysCDG -- it just won't be found by standard path finding algorithms, because it isn't acyclic (and there are potentially infinitely many cyclic paths in a graph).

Note that the definition of path signatures is irrelevant to the soundness of the cumulative procedure, only 
affecting its precision and performance: If a path is wrongfully considered to be resolved, then its reincarnation still appears in \texttt{violating\_paths} -- it just isn't recognized as the same. Therefore, in the worst case, a useless refinement is kept, worsening performance or (if to the point where the execution times out) leading to a false positive result of illegal information flow. If a path is thought to be recognized, even though it is actually a different path, then in the worst case a useful refinement is discarded, also possibly leading to a false positive result of illegal information flow.

However, for the non-cumulative procedure, it \emph{is} relevant to soundness that, if a common violating path exists (\ie one that represents the same actually possible information flow), then it must be found by the analysis. Otherwise, we could erroneously report secure information flow. Overapproximating common paths on the other hand is fine for soundness, with the same consequences for performance or false positives as with the cumulative procedure. We believe that our approach with the signature product graph and program locations as node signatures is sound, but currently do not provide formal assurances of that.

\section{Evaluation}\label{sec:evaluation}

\begin{table*}[tbp]
    \centering
    \caption{Results of running our analysis on some examples. The first line per examples denotes the numbers of explored states, the second denotes the runtimes of the analyses. A $\checkmark$ indicates correctly identified security, a $\times$ indicates a false positive. The values for information flow refinement are when combined with the ideal exploration refinement heuristic.}
    \label{tab:evaluation}
    \begin{tabular}{|l||c|c|c|c@{~~}c|c@{~~}c|c@{~~}c|}
    \hline
     & \multicolumn{3}{c|}{Exploration Refinement} & \multicolumn{6}{c|}{Information Flow Refinement} \\
    \cline{2-10}
    Example     & none    & PDs   & splitters & \multicolumn{2}{c|}{none} & \multicolumn{2}{c|}{cum.} & \multicolumn{2}{c|}{n.-cum.} \\
    \hline\hline
    Priorities  & 24      & 24    & 24     & 24    & \multirow{2}{*}{$\times$}     & 147   & \multirow{2}{*}{$\checkmark$} & 79    & \multirow{2}{*}{$\checkmark$} \\
                & 0.01s   & 0.02s & 0.02s  & 0.03s &                               & 0.13s &                               & 0.13s &                               \\
    \hline
    Many Slots     & 15      & 15    & 15     & 15    & \multirow{2}{*}{$\times$}     & 36   & \multirow{2}{*}{$\checkmark$} & 36    & \multirow{2}{*}{$\checkmark$} \\
                & 0.01s   & 0.02s & 0.02s  & 0.03s &                               & 0.01s &                               & 0.01s &                               \\
    \hline
    Arbiter     & 43      & 43    & 48     & 43    & \multirow{2}{*}{$\times$}     & 48   & \multirow{2}{*}{$\checkmark$} & 48    & \multirow{2}{*}{$\checkmark$} \\
                & 0.01s   & 0.02s & 0.03s  & 0.02s &                               & 0.04s &                               & 0.04s &                               \\
    \hline
    Transmitters & 3528620 & 170   & 170    & 170   & \multirow{2}{*}{$\checkmark$} & 170   & \multirow{2}{*}{$\checkmark$} & 170   & \multirow{2}{*}{$\checkmark$} \\
                & 130s    & 0.29s & 0.15s  & 0.19s &                               & 0.18s &                               & 0.18s &                               \\
    \hline
\end{tabular}
\end{table*}

To evaluate our approach, we have prototypically implemented it in Java for
SystemC, and ran it on a couple of case studies of shared bus architectures. The implementation and a full reproduction package are available as an artifact \cite{artifact}. The results are shown in \Cref{tab:evaluation}.

In the first case study (\emph{Priorities}), values with different priorities are communicated with different timing behaviors. This example does not require exploration refinement, but reports a false positive without information flow refinement.
The same holds for our second case study (\emph{Many Slots}), which models a bus with multiple slots where security levels are stored in addition to values.
In both case studies, the false positives are eliminated by both the cumulative and non-cumulative refinements. They are equally efficient, but the non-cumulative approach sometimes finds abstractions with fewer states.

In our third case study (\emph{Arbiter}), the bus is dynamically allocated.
Exploration refinement is not necessary here. The splitters heuristic even performs a slight overrefinement, however these additional states are required anyway to prove security, and are therefore also introduced by information flow refinement.

Our fourth case study (\emph{Transmitters}) has two transmitters of different security levels and with dynamic timing behavior.
Without exploration refinement, this case study suffers greatly from the state space explosion problem. Both our exploration heuristics reduce the state space dramatically, also bringing down the exploration time by a factor of $\approx 100$. Trying to construct the SysCDG without exploration refinement lead to an \texttt{OutOfMemoryError} on our machine\footnote{11th Gen Intel(R) i7-1165G7, 64GB RAM, Eclipse Temurin 25.0.3.9 JDK}.

The examples show that our heuristics can succeed in dramatically reducing state space and analysis time and in excluding spurious information flow, strengthening the efficiency and precision of the IFA. The partial descendants and splitters heuristics deliver similar results. Cumulative and non-cumulative refinement are similarly efficient, but non-cumulative can lead to fewer states.

\section{Related Work}\label{sec:related_work}

Abstraction refinement is common in model checking and abstract interpretation \cite{Clarke2000,Henzinger2002}.
\cite{Flanagan2003,Henzinger2003} refine abstractions using predicates relevant to shared state and inter-thread interference. While conceptually related to our splitters heuristic, their setting and refinement criteria are different and they do not target IFA. \cite{Livshits2009} uses probabilistic inference over constraints to infer specifications for information-flow problems, but is not geared towards embedded systems and only considers data flow.
\cite{Jhala2005} proposes path slicing to remove statements irrelevant to a counterexample trace before feasibility checking and refinement, which is related to our path-resolution heuristic in its use of dependence information for refinement, but targets safety rather than information flow.

Path conditions often serve a similar purpose as our information flow refinement, excluding spurious paths through dependence graphs by constructing SMT formulas capturing program executions and checking their satisfiability \cite{Snelting2006}. However, they become expensive or lose precision in the presence of loops, recursion, or concurrency, and are typically not geared towards embedded systems. \cite{Becker-Kupczok2024b} proposes to use timed path conditions for automatic security hardening of embedded systems, but does not provide for the derivation of such conditions. \cite{Mikulcak2018} uses timed path conditions for IFA in the signal-flow oriented Simulink language, but is not applicable to control-flow oriented languages like SystemC.

\cite{Taghdiri2010} refines path conditions automatically to improve the precision of an information-flow analysis. The approach is developed for Java bytecode and does not address embedded-system characteristics.
Since candidate witnesses are validated by concrete program execution, non-termination of the analyzed program can also prevent the analysis from reaching a conclusive result.

There are also numerous approaches for IFA on concurrent systems. Many of the recently developed ones are based on rely/guarantee reasoning \cite{Mantel2011,Schoepe2020} or separation logics \cite{Karbyshev2018,Eilers2023}, and as such require at least some manual effort with a high degree of expertise for most non-trivial cases.
Among other approaches, e.g. \cite{Gerking2018,Hecker2022}, they also often ensure the absence of timing leaks, though often in part by restricting high-dependent branching or synchronization.
Graph-based approaches \cite{Krinke2003,Giffhorn2015} can be fully automated, but often sacrifice precision or incur higher computational effort.
Also, none of these approaches account for the way timing-based synchronization, which is common in hardware designs, can rule out information flow, which is the main purpose of the approach from \cite{Becker-Kupczok2024a} that we build upon.
\cite{Nielson2017} provide information flow control for \textsc{Uppaal} timed automata, but their approach is specifically tailored to that formalism.

Hardware-level IFA has been addressed by gate-level and RTL information-flow tracking approaches \cite{Tiwari2009,Ardeshiricham2017}, including recent work on quantitative timing-sensitive leakage at RTL \cite{Reimann2024}. These approaches are complementary to ours, which operates on SystemC models and refines symbolic-execution abstractions rather than propagating hardware security labels.

Symbolic execution has been combined with IFA, e.g. in \cite{Phan2012,
Ryan2023}. However, these approaches do not take timing and concurrency into account. It has also been applied to embedded systems \cite{Cimatti2011,Le2013,Davidson2013}, but not
in the context of IFA.

To the best of our knowledge, abstraction refinement has not yet been applied to information flow analyses of embedded systems.

\section{Conclusion}\label{sec:conclusion}

In this paper, we presented an automated refinement approach for symbolic-execution-based information flow analysis of timed concurrent SystemC models. By dynamically adapting abstraction levels during the analysis, our approach substantially reduces the need for manually selected abstractions while preserving the precision required to capture timing- and control-dependent information flow. To this end, we introduced two exploration refinement heuristics, \emph{partial descendants} and \emph{splitters}, which reduce the symbolic state space by identifying branching decisions whose refinement is expected to do so. In addition, we proposed the \emph{path resolution} heuristic, which selectively refines the abstraction along paths containing potential illegal information flow to eliminate spurious flows introduced by overapproximation.

We implemented the proposed refinement techniques and evaluated them on several shared bus architectures. The experimental results demonstrate that the proposed heuristics automatically derive abstractions that enable efficient and precise information flow analysis, significantly reducing the manual effort required from the user while maintaining scalability. 

In future work, we plan to optimize the analysis performance, explore additional refinement heuristics, and scale the approach to handle larger, more complex industrial designs.

%
%
%
\bibliographystyle{splncs04}
\bibliography{bibliography}

\end{document}

%% file: preamble.tex
\usepackage{calc}
\usepackage{etoolbox}
\usepackage{mathtools}
\usepackage{amssymb}
\usepackage{dsfont}
\usepackage{bm}
\usepackage{cancel}
\usepackage{latexsym}
\usepackage{stmaryrd}

\AtBeginDocument{%
  \mathchardef\mathcomma\mathcode`\,
  \mathcode`\,="8000 
}
{\catcode`,=\active
  \gdef,{\mathcomma\discretionary{}{}{}}
}

\usepackage{ifthen}
\usepackage{multicol}
\usepackage{multirow}
\usepackage{tabulary}
\usepackage{longtable}
\usepackage{hhline}
\usepackage[labelfont=bf,labelsep=period]{caption}
\usepackage{subcaption}
\usepackage{placeins}
\usepackage{xcolor}
\usepackage{csquotes}
\usepackage[normalem]{ulem}
\usepackage{paralist}
\usepackage{enumitem}
\usepackage{algorithm}
\usepackage{algorithmicx}
\usepackage[commentColor=black]{algpseudocodex}

\algrenewcommand{\textproc}{\ttfamily}
\makeatletter
\newcommand{\MComment}[1]{%
  \algrenewcommand\alglinenumber[1]{\null}%
  \LComment{#1}
  \algrenewcommand\alglinenumber[1]{\footnotesize ##1:}%
}
\makeatother
\usepackage{listings}
\usepackage{xspace}
\xspaceremoveexception{-}
\makeatletter
\renewcommand*\@xspace@hook{%
    \ifx\@let@token-%
        \expandafter\@xspace@dash@i
    \fi
}
\def\@xspace@dash@i-{\futurelet\@let@token\@xspace@dash@ii}
\def\@xspace@dash@ii{%
    \ifx\@let@token-%
    \else
        \unskip
    \fi
    -%
}
\makeatother

\definecolor{darkgreen}{RGB}{0,128,0}
\definecolor{ctrlInstrColor}{RGB}{100,65,30}
\definecolor{waitInstrColor}{RGB}{0,90,40}
\definecolor{chanInstrColor}{RGB}{0,100,145}
\definecolor{ioInstrColor}{RGB}{0,0,128}
\definecolor{otherKeywordColor}{RGB}{110,45,175}
\definecolor{commentColor}{RGB}{65,65,65}
\definecolor{nameColor}{RGB}{0,0,0}

\lstdefinelanguage{SystemC}[]{C++}{			
    xleftmargin=2.5em,
    columns=flexible,
	alsoletter = <>,						
    commentstyle={\color{commentColor}},
	directivestyle={\color{red}}, 		     
	identifierstyle={\color{nameColor}},		
    keywords = [1]{while, if, else, switch, case, return, class},
    keywordstyle = [1]{\color{ctrlInstrColor}},
    keywords = [2]{const, friend, operator, struct, void, signed, unsigned, bool, char, int, long, float, double},
    keywordstyle = [2]{\color{otherKeywordColor}},
    keywords = [3]{true, false},
    keywordstyle = [3]{\color{otherKeywordColor}},
    keywords = [4]{wait, notify, SC_ZERO_TIME},
    keywordstyle = [4]{\color{waitInstrColor}},
    keywords = [5]{SC_MODULE, SC_CTOR},
    keywordstyle = [5]{\color{ctrlInstrColor}},
    keywords = [6]{read, write, sc_signal<int>, sc_signal<int, SC_MANY_WRITERS>},
    keywordstyle = [6]{\color{chanInstrColor}},
    keywords = [7]{sc_in<int>, sc_out<int>},
    keywordstyle = [7]{\color{chanInstrColor}},
}
\newcommand{\lstreset}{\lstset{basicstyle=\ttfamily\small,tabsize=4}}
\lstreset

\usepackage{tikz}
\usetikzlibrary{positioning}
\usetikzlibrary{fit}
\usetikzlibrary{calc}
\usetikzlibrary{matrix}
\usetikzlibrary{automata}
\usetikzlibrary{shapes}
\usetikzlibrary{arrows}
\usetikzlibrary{arrows.meta}
\usetikzlibrary{calc}
\usetikzlibrary{shapes}
\usetikzlibrary{decorations.pathreplacing}
\pgfdeclarelayer{bg}
\pgfsetlayers{bg,main}

\usepackage{hyperref}
\usepackage{cleveref}

\DeclareMathSymbol{\mlq}{\mathord}{operators}{'134}
\DeclareMathSymbol{\mrq}{\mathord}{operators}{'42}

\newcommand{\commentOut}[1]{}
\newcommand{\nmspace}{\relax\ifmmode\else\xspace\fi}
\newcommand{\mathMacro}[1]{\ensuremath{#1}\nmspace}

\newcommand{\mathEncSp}{\mathchoice{\mspace{-5mu}}{\mspace{-5mu}}{\mspace{-3mu}}{\mspace{-3mu}}}
\newcommand{\mathPreSp}{\mathchoice{\!}{\!}{}{}}
\newcommand{\flxqt}[1]{\relax\ifmmode{\mlq#1\mrq}\else{\enquote{#1}}\fi} 

\newcommand{\ie}{i.e.\xspace}

\newcommand{\eg}{e.g.\xspace}

\newcommand{\parens}[1]{\left(\vphantom{#1}\right.\mathEncSp#1\mathEncSp\left.\vphantom{#1}\right)}
\newcommand{\braces}[1]{\left\lbrace\vphantom{#1}\right.\mathEncSp#1\mathEncSp\left.\vphantom{#1}\right\rbrace}

\newcommand{\applyFunction}[2]{#1\mathPreSp\parens{#2}}

\newcommand{\set}[1]{\braces{#1}}
\newcommand{\with}{\mathrel{|}}
\newcommand{\union}{\mathbin{\cup}}
\newcommand{\bigUnion}{\bigcup}

\newcommand{\sign}{\mathMacro{\mathit{sign}}}
\newcommand{\signOf}[1]{\mathMacro{\applyFunction{\sign}{#1}}}

%% file: img/approach_24.tex
\begin{tikzpicture}[every node/.style={font={\LARGE}}]
\Large

\node[draw,rectangle,thick,label={[align=center]above:SystemC Design}] (scmodel) at (0,6) {\includegraphics[width=4cm]{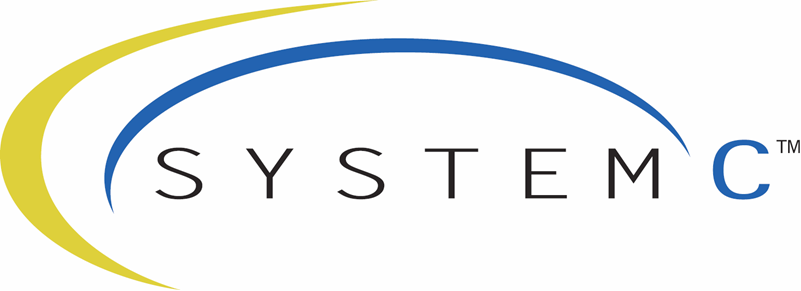}};

\node[draw,thick,rounded corners,align=center] (dpa) at (6,6) {Block-Wise\\Dependency Analysis};

\node[draw,thick,rectangle,label={[align=center]above:Program\\Dependence Graphs}] (pdgs) at (12,6) {\resizebox{4cm}{!}{\input{img/tiny_pdgs}}};

\node[draw,ultra thick,rounded corners,align=center] (syx) at (0,2.5) {Symbolic Execution\\with Abstract\\Interpretation};

\node[draw,ultra thick,rectangle,label={[align=center]below:Thread\\Interleaving Graph}] (tig) at (6,2.5) {\resizebox{4cm}{!}{\input{img/tiny_tig}}};

\node[draw,ultra thick,rounded corners,align=center] (isa) at (12,2.5) {Time and\\Event Sensitive\\Dependency Analysis};

\node[draw,ultra thick,rectangle,label={[align=center]below:SystemC\\Dependence Graph}] (sdg) at (18,2.5) {\resizebox{4cm}{!}{\input{img/tiny_sdg}}};

\node[draw,thick,rounded corners,align=center] (ifa) at (18,6) {Reachability\\Analysis};

\node[label={[align=center]above:Secure\\Information Flow}] (secure) at (23,6) {\includegraphics[width=1.5cm]{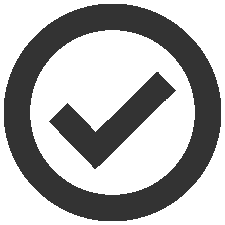}};

\node[draw,thick,rectangle,minimum size=2cm,label={[align=center]below:Violating\\Path}] (path) at (23,2.5) {\resizebox{1.75cm}{!}{\input{img/pathfinding_icon_tikz}}};

\draw[-Latex,thick] (scmodel) -- (dpa);
\draw[-Latex,thick] (dpa) -- (pdgs);
\draw[-Latex,thick] (pdgs) -- (isa);
\draw[-Latex,thick] (scmodel) -- (syx);
\draw[-Latex,thick] (syx) -- (tig);
\draw[-Latex,thick] (tig) -- (isa);
\draw[-Latex,thick] (isa) -- (sdg);
\draw[-Latex,thick] (sdg) -- (ifa);
\draw[-Latex,thick] (ifa) -- (secure);
\draw[-Latex,thick] (ifa) -- (path);

\end{tikzpicture}

%% file: img/tiny_pdgs.tex
\begin{tikzpicture}

\node[draw,rectangle,minimum width=1.5cm, minimum height=0.65cm] (a0) at (1.75,5) {};
\node[draw,circle,minimum size=1cm] (a1) at (0,4.5) {};
\node[draw,circle,minimum size=1cm] (a2) at (1,3.5) {};
\node[draw,circle,minimum size=1cm] (a3) at (2.5,3.5) {};
\node[draw,circle,minimum size=1cm] (a4) at (3.5,4.5) {};

\draw[-Latex] (a0) -- (a2);
\draw[-Latex] (a0) -- (a3);
\draw[dashed,-Latex] (a1) edge[bend left=15] (a3);
\draw[dashed,-Latex] (a2) edge[bend left=15] (a4);

\node[draw,rectangle,minimum width=1.5cm, minimum height=0.65cm] (b0) at (6.75,4.5) {};
\node[draw,circle,minimum size=1cm] (b1) at (5,4.0) {};
\node[draw,circle,minimum size=1cm] (b2) at (6.75,3.0) {};
\node[draw,circle,minimum size=1cm] (b3) at (6.0,1.5) {};
\node[draw,circle,minimum size=1cm] (b4) at (7.5,1.5) {};
\node[draw,circle,minimum size=1cm] (b5) at (8.5,4.0) {};

\draw[-Latex] (b0) -- (b2);
\draw[-Latex] (b2) -- (b3);
\draw[-Latex] (b2) -- (b4);
\draw[dashed,-Latex] (b1) -- (b2);
\draw[dashed,-Latex] (b3) edge[bend left] (b2);
\draw[dashed,-Latex] (b4) -- (b5);

\node[draw,rectangle,minimum width=1.5cm, minimum height=0.65cm] (c0) at (2.25,2.5) {};
\node[draw,circle,minimum size=1cm] (c1) at (0.5,2.0) {};
\node[draw,circle,minimum size=1cm] (c2) at (1.5,1.0) {};
\node[draw,circle,minimum size=1cm] (c3) at (3.0,1.0) {};
\node[draw,circle,minimum size=1cm] (c4) at (4.0,2.0) {};

\draw[-Latex] (c0) -- (c2);
\draw[-Latex] (c0) -- (c3);
\draw[dashed,-Latex] (c1) -- (c2);
\draw[dashed,-Latex] (c2) -- (c3);
\draw[dashed,-Latex] (c3) -- (c4);

\end{tikzpicture}

%% file: img/tiny_tig.tex
\begin{tikzpicture}[every node/.style={font={\normalsize}}]
\small

\node[draw,ellipse,minimum width=0.65cm, minimum height=1cm] (start) at (-0.5,3) {};
\node[draw,ellipse,minimum width=1cm,minimum height=0.65cm] (a) at (1,3.5) {};
\node[draw,ellipse,minimum width=1cm,minimum height=0.65cm] (b) at (1,2.5) {};
\node[draw,ellipse,minimum width=1cm,minimum height=0.65cm] (c) at (3,3.5) {};
\node[draw,ellipse,minimum width=1cm,minimum height=0.65cm] (d) at (3,2.5) {};
\node[draw,ellipse,minimum width=1cm,minimum height=0.65cm] (e) at (2,1.5) {};

\draw[-Latex] (start) -- (a) node[midway,above] {1};
\draw[-Latex] (start) -- (b) node[midway,below] {2};
\draw[-Latex] (a) -- (c) node[midway,above] {2};
\draw[-Latex] (b) -- (d) node[midway,below] {1};
\draw[-Latex] (c) -- (b) node[midway,below right=-1.0mm] {3};
\draw[-Latex] (d) -- (e) node[midway,below right] {3};
\draw[-Latex] (e) -- (b) node[midway,below left] {2};

\end{tikzpicture}

%% file: img/tiny_sdg.tex
\begin{tikzpicture}

\node[draw,rectangle,minimum width=1.5cm, minimum height=0.65cm] (a0) at (1.75,5) {};
\node[draw,circle,minimum size=1cm] (a1) at (0,4.5) {};
\node[draw,circle,minimum size=1cm] (a2) at (1,3.5) {};
\node[draw,circle,minimum size=1cm] (a3) at (2.5,3.5) {};
\node[draw,circle,minimum size=1cm] (a4) at (3.5,4.5) {};

\draw[-Latex] (a0) -- (a2);
\draw[-Latex] (a0) -- (a3);
\draw[dashed,-Latex] (a1) edge[bend left=15] (a3);
\draw[dashed,-Latex] (a2) edge[bend left=15] (a4);

\node[draw,rectangle,minimum width=1.5cm, minimum height=0.65cm] (b0) at (6.75,4.5) {};
\node[draw,circle,minimum size=1cm] (b1) at (5,4.0) {};
\node[draw,circle,minimum size=1cm] (b2) at (6.75,3.0) {};
\node[draw,circle,minimum size=1cm] (b3) at (6.0,1.5) {};
\node[draw,circle,minimum size=1cm] (b4) at (7.5,1.5) {};
\node[draw,circle,minimum size=1cm] (b5) at (8.5,4.0) {};

\node[draw,very thick,circle,minimum size=1cm] (b6) at (4.75,0.5) {};
\node[draw,very thick,circle,minimum size=1cm] (b7) at (5.4,2.55) {};

\draw[-Latex] (b0) -- (b2);
\draw[-Latex] (b2) -- (b3);
\draw[-Latex] (b2) -- (b4);
\draw[dashed,-Latex] (b1) -- (b2);
\draw[dashed,-Latex] (b3) -- (b6);
\draw[very thick,dash dot,-Latex] (b6) -- (b7);
\draw[dashed,-Latex] (b7) -- (b2);
\draw[dashed,-Latex] (b4) -- (b5);

\node[draw,rectangle,minimum width=1.5cm, minimum height=0.65cm] (c0) at (2.25,2.5) {};
\node[draw,circle,minimum size=1cm] (c1) at (0.5,2.0) {};
\node[draw,circle,minimum size=1cm] (c2) at (1.5,1.0) {};
\node[draw,circle,minimum size=1cm] (c3) at (3.0,1.0) {};
\node[draw,circle,minimum size=1cm] (c4) at (4.0,2.0) {};

\draw[-Latex] (c0) -- (c2);
\draw[-Latex] (c0) -- (c3);
\draw[dashed,-Latex] (c1) -- (c2);
\draw[dashed,-Latex] (c2) -- (c3);
\draw[dashed,-Latex] (c3) -- (c4);

\draw[very thick] (a3) edge[bend right=15] (5.0,3.25);
\draw[very thick] (5.0,3.25) edge[bend right=15,-Latex] (b0);
\draw[very thick,-Latex] (b3) edge[bend left] (c0);
\draw[dashed] (b6) edge[bend left=17.5] (1.25,0);
\draw[dashed] (1.25,0) edge[bend left=40,-Latex] (c1);
\draw[dashed,-Latex] (c4) edge[bend left=0] (b7);

\end{tikzpicture}

%% file: img/pathfinding_icon_tikz.tex
\begin{tikzpicture}[every node/.style={draw,line width=2.5mm,circle}, emph/.style={draw,line width=2mm,-{Latex[length=12mm,width=10mm]}}, norm/.style={draw,line width=0.5mm,-{Latex[length=6mm,width=5mm]}}]
    \node[draw,circle] (A) at (0,1) {};
    \node[draw,circle] (B) at (0.5,5) {};
    \node[draw,circle] (C) at (5.75,2) {};
    \node[draw,circle] (D) at (8,5.5) {};
    \node[draw,circle] (E) at (8,-2) {};
    \node[draw,circle] (F) at (11,2) {};
    \node[draw,circle] (G) at (3.0,-2.5) {};

    \draw[emph] (A) edge (C);
    \draw[emph] (C) edge (D);
    \draw[emph] (D) edge (F);
    \draw[emph] (F) edge (E);
\end{tikzpicture}

%% file: img/approach.tex
\begin{tikzpicture}[every node/.style={font={\LARGE},align=center},
        rectn/.style 2 args={inner sep=0pt, outer sep=0pt, fit={(#1)(#2)}}]
    \Large

    \node[draw,thick,rounded corners,rectn={3.5,8}{7,3}] (execution) {Symbolic\\Execution};

    \node[draw,thick,rounded corners,rectn={11,8}{14.5,3}] (refinement) {};
    \draw[thick] (refinement.east) -- (refinement.west);

    \node[draw=none] (exploration_refinement) at ($(refinement.north)!0.25!(refinement.south)$) {Exploration\\Refinement};
    
    \node[draw=none] (path_refinement) at ($(refinement.north)!0.75!(refinement.south)$) {Information\\Flow\\Refinement};

    \node[draw,thick,rectangle,label={[name=abstraction_label]above:Abstraction}] (abstraction) at (9,4.5) {$\cancel{a},\ \cancel{b},\ c,$\\[-0.3\baselineskip]$x,\ \cancel{y},\ z$};
    
    \node[draw,thick,rectangle,label={above:SystemC Design}] (scmodel) at (0,0|-execution.center) {\includegraphics[width=4cm]{img/systemc}};
    
    \node[draw,thick,rectangle,label={below:Information\\Flow Policy}] (policy) at (0,-0.25) {\resizebox{4cm}{!}{\input{img/policy}}};

    \node[draw,thick,rounded corners,rectn={execution.west|-policy.north}{refinement.east|-policy.south}] (analysis) {};
    \node[draw=none] at (analysis.center) {Information\\Flow Analysis};

    \node[label={below:Secure\\Information Flow}] (secure) at (18,-0.25) {\includegraphics[width=1.5cm]{img/secure_icon_3}};

    \node[label={above:Possible\\Violation}] (insecure) at (path_refinement-|secure) {\includegraphics[width=1.5cm]{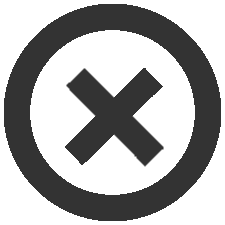}};

    \draw[-Latex,thick] (scmodel) -- (execution);
    \draw[-Latex,thick] (scmodel) -- (analysis.north west);
    \draw[-Latex,thick] (policy) -- (analysis);
    \draw[-Latex,thick] ([yshift=-0.25cm]execution.north east) -- node[above,midway] {branch} ([yshift=-0.25cm]refinement.north west);
    \draw[-Latex,thick] ([yshift=-1cm]refinement.north west) -- node[above,midway] {continue} ([yshift=-1cm]execution.north east);
    \draw[-Latex,thick] ([yshift=-1.75cm]refinement.north west) -- coordinate[midway] (exploration_restart) node[above,midway] {restart} ([yshift=-1.75cm]execution.north east);
    \draw[-Latex,thick] ([yshift=0.25cm]refinement.south west) -- coordinate[midway] (path_restart) node[above,below] {restart} ([yshift=0.25cm]execution.south east);
    \draw[-Latex,thick] (execution) -- node[pos=0.625,right] {exploration\\successful} coordinate[midway] (tig_coordinate) (analysis.north-|execution);
    \draw[-Latex,thick] (analysis) -- (secure);
    \draw[-Latex,thick] (analysis.north-|refinement) -- coordinate[midway] (analysis_path) node[pos=0.375,left] {illegal path\\detected} (refinement);
    \draw[-Latex,thick] (refinement.east|-path_refinement) -- coordinate[midway] (refinement_path) (insecure);
    
    \node[draw,thick,rectangle,minimum size=2cm,label={[name=path_label]right:SysCDG\\Path}] (path) at (analysis_path-|refinement_path) {\resizebox{1.75cm}{!}{\input{img/pathfinding_icon_tikz}}};

    \draw[dotted,thick] ([yshift=-0.25cm]abstraction_label.north) -- (exploration_restart);
    \draw[dotted,thick] (abstraction) -- (path_restart);
    \draw[dotted,thick] (analysis_path) -- (path);
    \draw[dotted,thick] (refinement_path) -- (path);
\end{tikzpicture}

%% file: img/path_resolution_sender.tex
\begin{lstlisting}[escapeinside={§}{§}]
while (true) {
  flag = §\tikzmark{flag_set}§1 - flag;
  if (fl§\tikzmark{flag_check}§ag == 0) {
    val = low_in;
    low_ev.noti§\tikzmark{notify}§fy(1, SC_MS);
  } else {
    val = §\tikzmark{high_write}§high_in;
    high_ev.notify(1, SC_MS);
  }
  wait(2, SC_MS);
}
\end{lstlisting}

%% file: img/resolution_data_of_control.tex
\begin{lstlisting}[showlines=true,escapeinside={§}{§}]
wait(1, SC_MS);
while(true) {
  if (flag§\tikzmark{flag_check_b}§ == 0)
    low_out§\tikzmark{low_read_b}§ = val;
  else
    high_out = val;
  wait(2, SC_MS)
}

\end{lstlisting}

%% file: img/resolution_slice_of_control.tex
\begin{lstlisting}[showlines=true,escapeinside={§}{§}]
while(true) {
  wait(lo§\tikzmark{wait_c}§w_ev);
  low_out§\tikzmark{low_read_c}§ = val;
}

\end{lstlisting}

%% file: img/resolution_slice_of_slice.tex
\begin{lstlisting}[escapeinside={§}{§}]
wait(1, SC_MS);
while (true) {
  data = §\tikzmark{high_read_d}§2 * val;
  if (fla§\tikzmark{flag_check_d}§g != 0)  data §\tikzmark{overwrite_d}§= 5;
  low_out §\tikzmark{low_write_d}§= data;
  wait(2, SC_MS);
}
\end{lstlisting}